\documentclass[12pt,preprint]{aastex}

\shorttitle{SSSPM J1549$-$3544 IS NOT A WHITE DWARF}
\shortauthors{J. Farihi}

\begin{document}

\title{SSSPM J1549$-$3544 IS NOT A WHITE DWARF}

\author{J. Farihi\altaffilmark{1,2},
	P. R. Wood\altaffilmark{3}, \&
	B. Stalder\altaffilmark{4}}

\altaffiltext{1}{Gemini Observatory,
			Northern Operations,
			670 North A'ohoku Place,
			Hilo, HI 96720; jfarihi@gemini.edu}
\altaffiltext{2}{Department of Physics \& Astronomy,
			University of California,
			Los Angeles, CA 90095}
\altaffiltext{3}{Research School of Astronomy \& Astrophysics,
			Australian National University,
			Cotter Road, Weston Creek ACT 2611,
			Australia; wood@mso.anu.edu.au}
\altaffiltext{4}{Institute for Astronomy,
			University of Hawaii,
			2680 Woodlawn Drive,
			Honolulu, HI 96822; bstalder@ifa.hawaii.edu}

\begin{abstract}

Spectroscopy and photometry demonstrate that
SSSPM J1549$-$3544 is not a cool white dwarf
but a high velocity metal-poor halo star.

\end{abstract}


\section{INTRODUCTION}

Cool white dwarfs are typically identified through
the combination of proper motion and apparent magnitude
called reduced proper motion \citep{lie79,giz97b,sal02,
sal03,lep05}.  At a given brightness and color, white
dwarfs will have larger proper motions than main sequence
stars owing to closer distances from the Earth implied
by their small radii.

Metal-poor stars, or subdwarfs, also stand out well
in reduced proper motion diagrams due to their lower
luminosities (for the same color class) and typically
higher velocities.  Subluminous main sequence stars which
distinguish themselves in reduced proper motion are kinematic
members of the thick disk and halo \citep{mih81,bin87,bin98}.
Subdwarfs have long been a contaminant in surveys aiming
to find cool white dwarfs because both stellar types are
mixed in a reduced proper motion diagram at intermediate
colors.  Cool white dwarfs must be confirmed spectroscopically
\citep{lie79,giz97b,sal02,far04,kil04}.

Current complete and ongoing large-scale proper motion
surveys are discovering numerous exotic objects including
extreme subdwarfs, brown dwarfs, and cool white dwarfs
(see \citet{opp01,sch04,lep05} and references therein).
Most if not all of these searches are utilizing available
large-scale optical photographic sky survey catalogs.
This is potentially problematic for the identification
of cool white dwarfs because optical photographic
bandpasses do not yield broad color baselines, and
photographic photometry has intrinsically large errors.
These facts lead to a large overlap in the white dwarf
and subdwarf sequences in reduced proper motion diagrams
\citep{sal02}.  Yet even with spectroscopy, subdwarfs and
other weak-lined objects can appear featureless in low
resolution and/or low signal-to-noise (S/N) measurements --
potentially mistaken as DC white dwarfs \citep{mcc99,
far04}.  Identification of degenerates with $T_{\rm eff}=
4000-5000$ K should include good S/N spectroscopy covering
blue-green optical wavelengths \citep{far04,kil04}.

Contrary to the claim of \citet{sch04}, this paper
presents evidence that the high proper motion star
SSSPM J1549$-$3544 ($15^{\rm h}$ $48^{\rm m}$ $40.23
^{\rm s}$, $-35\arcdeg$ $44'$ $25.5''$, J2000) is neither
the nearest cool white dwarf nor a degenerate star, but
rather a very high velocity halo star passing through the
solar neighborhood.

\section{OBSERVATIONS \& DATA}

Optical $BVRI$ photometric data were obtained 14
February 2005 with the Orthogonal Parallel Transfer
Imaging Camera (OPTIC; \citealt{ton02}) on the University
of Hawaii 2.2 meter telescope at Mauna Kea.  Observing 
conditions were photometric with seeing $\sim1.0''$ and
some gusty winds $20-40$ mph.  Three identical exposures
of SSSPM J1549$-$3544 were taken at each bandpass for a
total integration time of $12-30$ seconds, yielding S/N
$>10$ at all wavelengths in a $0.7''$ aperture radius.
A standard Landolt field \citep{lan92} was observed
similarly immediately following the science target.

The images were flat fielded, cleaned of bad pixels
in the region of interest, and combined into a single
frame at each wavelength for photometry.  Calibration
was performed with three stars and aperture photometry
was executed using standard IRAF packages.  Calibrator
stars were measured in a $1.4''$ aperture radius and
the science target data were corrected to this standard
aperture and adjusted for relative extinction.  The
results are listed in Table \ref{tbl-1}.

Optical $4200-10000$ \AA \ spectroscopic data were
obtained 18 March 2005 with the Double Beam Spectrograph
(DBS; \citealt{rod88}) on the Australian National University
2.3m Telescope at Siding Spring Observatory.  Only the red
arm of the spectrograph was operated, with no dichroic in the
beam.  The 158 l/mm grating used gave a spectral resolution of
10 \AA \ (2.5 pixels) with the $2''$ slit.  The seeing was $\sim
2.3''$ and the sky was clear.

\begin{figure}

\plotone{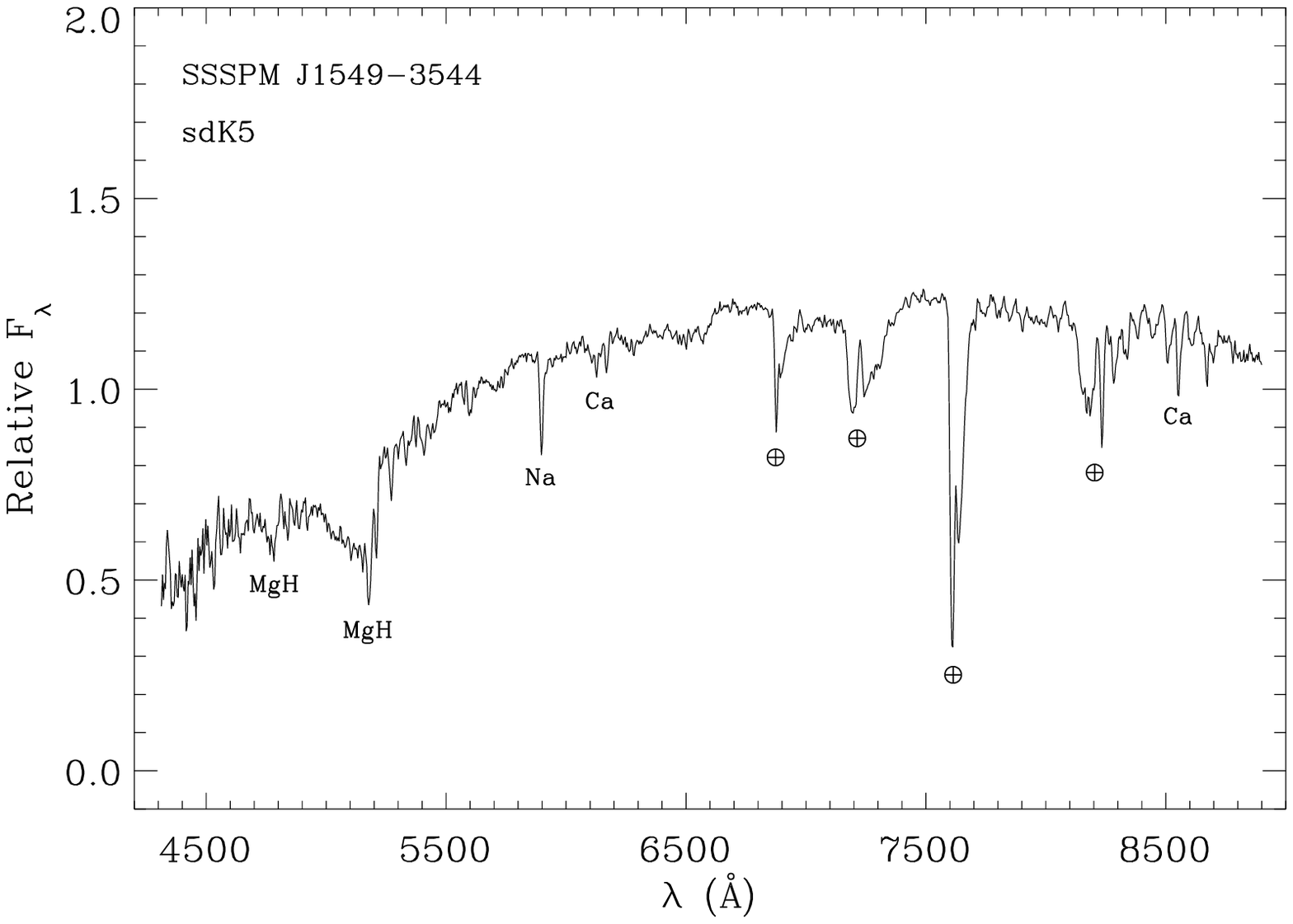}
\caption{Optical spectrum of SSSPM J1549$-$3544 taken with
the Double Beam Spectrograph on the 2.3 meter telescope at
Siding Springs Observatory.  The data are flux calibrated
and normalized near 5500\,\AA{}.  The most prominent stellar
absorption features are labelled along with telluric O$_2$
and H$_2$O bands.  The spectral type should be considered
preliminary (\S3).
\label{fig1}}
\end{figure}

SSSPM J1549$-$3544 and the flux calibrator LTT 6248
were each observed for 600 seconds, yielding S/N $>20$
over the majority of the science target wavelength shown
in Figure \ref{fig1}.  The spectral images were flat fielded,
cleaned of bad pixels in the region of interest, and flux
calibrated spectra were extracted using standard IRAF packages.
The spectra were trimmed to show the region of interest, while
no attempt was made to remove telluric features.

\section{RESULTS}

The combined optical and near-infrared colors of SSSPM
J1549$-$3544 are consistent with a main sequence star of
$T_{\rm eff}\sim4200$ K, late K spectral type \citep{bes88}.
The near-infrared magnitudes are too bright for known cool
and ultracool white dwarfs \citep{far05}.  For example, the
reddest known degenerates have colors which peak near $I-K\la
1.1$, after which they become bluer again as shown by observation
and models (excepting pure helium models at $T_{\rm eff}\la4250$
K; \citealt{ber95,ber01}), while SSSPM J1549$-$3544 has $I-K_s=1.58$.

The spectrum in Figure \ref{fig1} confirms the status of
SSSPM J1549$-$3544 as a nondegenerate star, and the deep
MgH feature near 5200 \AA \ indicates it is metal-poor
\citep{rei00}.  Other unlabelled spectral features present
include: weak H$\alpha$, several Fe and Ca lines, plus some
weak CaH and TiO.  These latter features are difficult to
distinguish as they are located, for the most part, within
and around the prominent telluric bands \citep{kir91} which
are also seen in the calibrator star spectrum.  

Following the methodology of \citet{giz97a}, a spectral
type and subdwarf class of SSSPM J1549$-$3544 was estimated 
by measuring the TiO5, CaH1, CaH2, \& CaH3 spectroscopic indices,
which are listed in Table \ref{tbl-2}.  Using equations $1-3$ in
\citet{giz97a}, a spectral type of K5 is found consistently across 
all three relations, and type sdK5 is found using equation 7 in
the same work.  However, given the fact that the spectral resolution
achieved here is $\sim3$ times lower than that of \citet{giz97a}, 
the spectroscopic measurments may be unreliable but are presented
here as a rough guide.  There are no distinctions between sdK and
esdK stars before spectral type K7, and, combined with the uncertain
spectroscopy, it is difficult to say whether or not SSSPM J1549$-$3544
is an extreme subdwarf without a direct metallicity measurment.  The
MgH band near 5200 \AA \ is quite strong and an extreme subdwarf
classification might be appropriate \citep{rei00}.

In order to assess the kinematics of SSSPM J1549$-$3544,
a distance must be estimated.  Conservatively assuming this
star lies two magnitudes below the main sequence at spectral
type K5, it would have $M_V\approx9.5$ mag \citep{dri00,rei00}.
This would place the star at $d=114$ pc with a tangential speed
of 430 km s$^{-1}$ based on the astrometic data in \citet{sch04}.
Furthermore, fourteen spectral lines (many without high S/N)
were used to measure the radial velocity, the crude result
being $v_r=+210\pm70$ km s$^{-1}$, corrected for the motion of
the Earth along the line of sight on the date of observation.
Combining all this data, a Galactic $UVW$ space motion was
calculated, corrected for the solar motion, $(U,V,W)=(-9,+12
+7)$ km s$^{-1}$ \citep{mih81}, relative to the local standard
of rest ($U$ positive toward the Galactic anticenter, $V$ positive
in the direction of Galactic rotation, $W$ positive toward the North
Galactic Pole).  The result is $(U,V,W)=(-46,-465,+42)$ km s$^{-1}$,
making SSSPM J1549$-$3544 a halo star \citep{jah97,bee00}.  If we
instead assume $v_r=0$ due to the potential unreliability of the
measured radial velocity, the result becomes $(U,V,W)=(143,-392,-10)$
km s$^{-1}$, which is still consistent with halo membership.  Despite
a fairly conservative assumption, the 114 pc distance estimate and
space motions should be considered preliminary as subdwarfs can span
a wide range of absolute magnitudes.  If SSSPM J1549$-$3544 is closer
to 150 pc, or around 1.5 magnitudes below the main sequence, it would
have a total heliocentric velocity in the range $550-590$ km s$^{-1}$.
At one magnitude below the main sequence, or 180 pc, SSSPM J1549$-$3544
would have a total Galactocentric velocity of $\sim470$ km s$^{-1}$
and would be among the fastest moving stars ever seen
\citep{car88,car96}.

\section{CONCLUSION}

The star SSSPM J1549$-$3544 is shown to be a metal-poor
sdK star rather than a cool white dwarf as previously claimed
\citep{sch04}.  It is possible this star is an extreme K subdwarf,
but in any case it is a very high velocity halo star passing through
the solar neighborhood.  With many new proper motion objects being
discovered and studied, it is critical to correctly distinguish cool
white dwarfs from subdwarf contaminants.  There is great science
potential in the oldest degenerates.

\acknowledgments

Some data used in this paper are part of the Two Micron All
Sky Survey, a joint project of the University of Massachusetts
and the Infrared Processing and Analysis Center / California
Institute of Technology, funded by the National Aeronautics
and Space Administration (NASA) and the National Science
Foundation.  J. Farihi acknowledges support by grants from
NASA.  PRW acknowledges the support provided by a grant
from the Australian Research Council.

\clearpage

\clearpage

\begin{deluxetable}{lcc}
%\tabletypesize{\small}
\tablecaption{Optical \& Near Infrared Photometric Data\label{tbl-1}}
\tablewidth{0pt}
\tablehead{
\colhead{Band} 			&
\colhead{$\lambda_{0}$ ($\mu$m)}&
\colhead{SSSPM J1549$-$3544 (mag)}}

\startdata

$B$		&0.44	&16.13\\
$V$		&0.55	&14.78\\
$R$ 		&0.64	&14.00\\
$I$ 		&0.80 	&13.20\\
$J$		&1.25	&12.34\\
$H$ 		&1.63	&11.77\\
$K_s$		&2.16	&11.62\\

\enddata

\tablecomments{The uncertainties are all $\leq$ 5\%.
$JHK_s$ data are taken from 2MASS \citep{cut03}.}

\end{deluxetable}

\clearpage

\begin{deluxetable}{lcc}
%\tabletypesize{\small}
\tablecaption{Spectroscopic Measurements\label{tbl-2}}
\tablewidth{0pt}
\tablehead{
\colhead{Band} 		&
\colhead{Strength}}

\startdata

TiO5		&0.986\\
CaH1		&0.983\\
CaH2		&1.034\\
CaH3		&0.997\\

\enddata

\tablecomments{Spectroscopic Band Measurements of SSSPM
J1549$-$3544, using the method of \citet{giz97a}.  A spectral
type of sdK5 is found on the basis of these data.}

\end{deluxetable}


\begin{thebibliography}{}

\bibitem[Beers et al.(2000)]{bee00} Beers, T., Chiba, M., Yoshii,
	Y., Platais, I., Hanson, R., Fuchs, B., \& Rossi, S. 2000,
	\aj, 119, 2866

\bibitem[Bergeron et al.(2001)]{ber01} Bergeron, P., Leggett, S., \&
	Ruiz, M. 2001, \apjs, 133, 413

\bibitem[Bergeron et al.(1995)]{ber95} Bergeron, P., Wesemael, F., \&
	Beauchamp, A. 1995, \pasp, 107, 1047

\bibitem[Bessell \& Brett(1988)]{bes88} Bessell, M. S., \& Brett,
	J. M. 1988, \pasp, 100, 1134

\bibitem[Binney \& Merrifield(1998)]{bin98} Binney, J., \& Merrifield,
	M. 1998, in Galactic Astronomy, (New Jersey: Princeton)

\bibitem[Binney \& Tremaine(1987)]{bin87} Binney, J., \& Tremaine, S.
	1987, in Galactic Dynamics, (New Jersey: Princeton)

\bibitem[Carney et al.(1988)]{car88} Carney, B. W., Laird, J. B., Latham,
	D. W. 1988, \aj, 96, 560
	
\bibitem[Carney et al.(1996)]{car96} Carney, B. W., Laird, J. B., Latham,
	D. W., \& Aguilar, L. A. 1996, \aj, 112, 668

\bibitem[Cutri et al.(2003)]{cut03} Cutri, R., et al. 2003, 2MASS All
	Sky Catalog of Point Sources (IPAC/CIT)
	
\bibitem[Farihi(2004)]{far04} Farihi, J. 2004b, Ph.D. Thesis,
	UCLA\footnote{Available at http://www.whitedwarf.org}
	
\bibitem[Farihi(2005)]{far05} Farihi, J. 2005, \aj, 129, 2382

\bibitem[Gizis(1997)]{giz97a} Gizis, J. E. 1997, \aj, 113, 806

\bibitem[Gizis \& Reid(1997)]{giz97b} Gizis, J. E., \& Reid, I. N.
	1997, \pasp, 109, 849
	
\bibitem[Jahrei\ss \ \& Wielen(1997)]{jah97} Jahrei\ss, H., \& Wielen, R.
	1997, Hipparcos '97, ed. B. Battrick (Noordwijk: ESA SP-402), 675
	
\bibitem[Drilling \& Landolt(2000)]{dri00} Drilling, J. S., \& Landolt,
	A. U. 2000, in Allen's Astrophysical Quantities, 4$^{\rm th}$
	edition, ed. A. N. Cox (New York: Springer-Verlag), 388
	
\bibitem[Kirkpatrick et al.(1991)]{kir91} Kirkpatrick, J., Henry,
	T., \& McCarthy, D. 1991, \apjs, 77, 417 

\bibitem[Landolt(1992)]{lan92} Landolt, A. U. 1992, \aj, 104, 340
	 
\bibitem[L\'epine \& Shara(2005)]{lep05} L\'epine. S. \& Shara, M.
	M. 2005, \aj, 1483
	
\bibitem[Liebert et al.(1979)]{lie79} Liebert, J., Dahn, C. C., Gresham,
	M., Strittmatter, P. A. 1979, \apj, 233, 226

\bibitem[Kilic et al.(2004)]{kil04} Kilic, M., Winget, D. E., von
	Hippel, T., \& Claver, C. F. 2004, \aj, 128, 1825

\bibitem[McCook \& Sion(1999)]{mcc99} McCook, G., \& Sion, E. 1999,
	\apjs, 121, 1

\bibitem[Mihalas \& Binney(1981)]{mih81} Mihalas, D., \& Binney, J.
	1981, in Galactic Astronomy, (San Francisco: W.  H.  Freeman
	\& Co.)		

\bibitem[Oppenheimer et al.(2001)]{opp01} Oppenheimer, B. R. Hambly,
	N. C., Digby, A. P., Hodgkin, S. T., \& Saumon, D. 2001,
	Science, 292, 698
	
\bibitem[Reid \& Hawley(2000)]{rei00} Reid, I., \& Hawley, S. 2000,
	in New Light on Dark Stars, (New York: Springer)
	
\bibitem[Rodgers et al.(1988)]{rod88} Rodgers, A. W., Conroy, P., \&
	Bloxham, G. 1988, \pasp, 100, 626
	
\bibitem[Salim \& Gould(2002)]{sal02} Salim, S., \& Gould, A. 2002,
	\apj, 575, 83

\bibitem[Salim \& Gould(2003)]{sal03} Salim, S., \& Gould, A. 2003,
	\apj, 582, 1011

\bibitem[Scholz et al.(2004)]{sch04} Scholz, R., Lehmann, I., Matute,
	I., \& Zinnecker, H. 2004, \aap, 425, 519

\bibitem[Tonry et al.(2002)]{ton02} Tonry, J. L., Luppino, G. A.,
	Kaiser, N., Burke, B. E., \& Jacoby, G. H. 2002, SPIE,
	4836, 206    
	
	
\end{thebibliography}
\end{document}